# A generalized enhanced Fourier law and underlying connections to major frameworks for quasi-ballistic phonon transport

**Ashok T. Ramu**[1]
Department of Electrical and Computer Engineering
University of California Santa Barbara
Santa Barbara, CA 93106-9560
ashok.ramu@gmail.com

**John E. Bowers**
Department of Electrical and Computer Engineering
University of California Santa Barbara
Santa Barbara, CA 93106-9560
bowers@ece.ucsb.edu

**ABSTRACT**

*An enhanced Fourier law (EFL) that accounts for quasi-ballistic phonon transport effects in a formulation entirely in terms of physical observables, is derived from the Boltzmann transport equation, assuming a gray population of quasi-ballistic phonon modes. This equation is generalized to an arbitrary phonon population. Other phonon transport models are analyzed in the context of the generalized EFL and connections are made between the generalized EFL and other models, revealing the essential unity of seemingly disparate models reported in the literature. A novel experiment is suggested to extract the accumulation function of bulk matter, and analyzed using the generalized EFL developed in this work.*

## 1. INTRODUCTION

Reports of significant room-temperature quasi-ballistic phonon transport [1-3] have spurred modeling efforts [4-11] aimed at explaining observed experimental results and predicting new effects. Each approach has specific advantages in certain situations, resulting in a wide variety of mathematical formulations. The aim of this article is to derive and generalize a useful,

---

[1] Corresponding author email: ashok.ramu@gmail.com

Fourier-like formulation, which we term the enhanced Fourier law, and explore its correspondence with other major formulations reported in the literature.

Few analytical or semi-analytical phonon conduction models exist that treat the full non-linear Boltzmann transport equation (BTE) [12]. Fully numerical solutions are outside the purview of this paper. All models discussed here assume a reference temperature and hence a reference equilibrium Bose distribution function, and assume the validity of the linearized BTE under the relaxation-time approximation.

We classify heat transfer models as gray models, two-channel models and quasi-full spectrum models. Grey models assume a constant velocity and lifetime for all phonon modes, and form the lowest in the hierarchy of heat-transfer models. Two-channel models, as the name suggests, divide the phonon spectrum into two conducting channels, one or both of which are assumed to follow the Fourier law with possibly different reference temperatures. The cut-off between the two channels is assumed arbitrary, which renders questionable the generality of these models. Quasi-full spectral models assume a single temperature, to which all phonon modes tend to equilibrate. This temperature may either be that of a background reservoir of phonons [4], or may be explicitly solved [5]. We begin by briefly summarizing the models that we will be examining.

## 2. SUMMARY OF MAJOR QUASI-BALLISTIC TRANSPORT FRAMEWORKS

### 2A. The enhanced Fourier law [8]

Ramu *et al.* proposed a technique to arrive at the heat-flux of a quasi-ballistic mode directly from the BTE by truncating the spherical harmonic expansion of the distribution function at the *l*=2 order in angular momentum. Since the heat-flux is a physically more accessible quantity than the distribution function, this formalism has certain advantages over others, some of which are (a) the angular integrals over distribution functions have already been performed, and non-locality of the quasi-ballistic heat-flux emerges naturally, (b) energy conservation is easier to enforce, (c) modal suppression functions may easily be derived, as exemplified in Sec. 4, and (d) non-Fourier laws of any desired level of accuracy may be easily obtained, as outlined in Sec. 4.

In its original form the EFL was a two-channel model [8]. In this work, we present a generalized EFL that should be classified as a quasi-full spectral BTE solution.

## 2B. Chen's ballistic-diffusive equations [6] and Miranda *et al.*'s series solution [7]

These two models are quasi-full spectral models. Despite the fact that on the surface, these seem like two-channel models, it must be recognized that in both cases, the overall BTE solution is simply written as a sum of a homogenous solution and a particular solution, which gives the appearance of a two-channel model. However, only one reference temperature is invoked for the diffusive part. Chen's ballistic diffusive equations (BDE) and Miranda *et al.* recognize that the homogenous Boltzmann equation is a damped advection equation, to which a closed-form solution exists. For the inhomogenous solution, while the BDE assumes the Fourier law, the solution of Miranda *et al.* assumes a gray medium (constant relaxation time for all modes) and expands the BTE solution in a Taylor series in the spatial coordinate to arrive at a beyond-Fourier constitutive law for the quasi-ballistic heat-flux.

## 2C. Weakly quasi-ballistic solution of Maznev *et al.* [4] and generalized BTE solution of Hua and Minnich [5]

In the context of the transient grating experiment, Maznev *et al.* presented a quasi-full spectral BTE solution by taking the Fourier transform of the BTE and writing the set of coupled BTEs for each mode as an eigen-value equation, which was solved for the transient grating decay time. By comparing the decay time with the Fourier law prediction, a correction factor called the "suppression function" was derived, from which the mean-free path accumulation function [13] could be recovered via a reconstruction technique [14]. Although their discussion begins with a two-channel BTE, Maznev *et al.* find that the Fourier law emerges as a limiting case of their quasi-ballistic equations, and seamlessly "merge" the two channels.

Hua and Minnich subsequently solved the full BTE semi-analytically with an improved energy conservation equation that takes into account the non-equilibrium modes instead of the usual procedure of ignoring them and lumping the equilibrium modes into a heat-capacity term. They found that when the ballistic decay time is much smaller than the overall thermal decay time, their suppression function reduced to that of Maznev *et al.* They called this the "weakly quasi-ballistic limit". On the other hand, when the two characteristic times were comparable, the grating decay rate was no longer exponential and a new transport regime emerged, one that could not be described by an effective mean-free-path dependent Fourier law. They called this the "strongly quasi-ballistic limit". These are of course in addition to the more familiar purely diffusive and purely ballistic limits of heat conduction, which the solution of Hua and Minnich could also reproduce. They derived a new suppression function for the transient grating experiment that was valid in all regimes of phonon conduction.

By assuming the modal decay time to be much less than the thermal time constant for decay of the transient grating, Maznev *et al.*'s solution automatically corresponds to the weakly quasi-ballistic limit of Hua and Minnich.

## 2D. The models of Regner *et al.* [25] and Yang and Dames [26]

Regner *et al.* [25] have proposed a gray model (constant mean-free path and constant velocity for all modes). They truncate the spherical harmonic approximation of the distribution function at the lowest non-trivial order, namely $l$=1 as opposed to $l$=2 of the enhanced Fourier law. In fact, except for different boundary conditions, their formulation is essentially the Cattaneo equation, as can be seen by inserting the equilibrium Bose distribution in Eq. (7a) of that work and summing over all modes. However, two counter-propagating heat-fluxes are separately solved for, which compensates for the low- order to which each is analyzed. The grey assumption makes it difficult to compare with other approaches.

Yang and Dames [26] have also used the Milne-Eddington approximation; however they have generalized their treatment to a non-gray population. Here again, introduction of forward and reverse propagating heat-fluxes compensates for the low order of spherical harmonic expansion. But for the way that the anisotropy of phonon population in $k$-space is handled, the approach of Yang and Dames is similar to the EFL. Their formulation has one advantage, namely that boundary conditions may be easily prescribed for one planar surface; however having two surfaces or extension to three dimensions engenders multiple phonon reflections, and it is not clear how this can be handled by their formulation. Solution of the EFL equations with general non-white partially specular boundaries in three dimensions also remains unclear at the time of this writing.

## 3. THE GENERALIZED ENHANCED FOURIER LAW

The enhanced Fourier law (EFL), as presented in [8] handles quasi-ballistic modes of a single mean-free path. It has utility in explaining the transient grating [15] and the frequency-domain thermoreflectance experiments [10,11]. Here we generalize it to handle arbitrary mean free path (MFP) spectra. To do so, we track the original derivation of [8] until the assumption of constant mean-free path is made. The generalization is necessary for meaningful comparison with the equations of Maznev *et al.* who used a Fourier transform in both time and space to reduce the set of BTEs for the phonon modes to an eigen-value equation. The following discussion additionally derives a generalized EFL, a differential equation for heat transfer by modes of multiple MFPs. This is of vital importance to realistic thermal modeling of silicon devices, because crystalline silicon has a wide spectrum, with phonons of about three orders of magnitudes in MFPs contributing significantly to thermal transport. Although in this section we will be focusing on 1D transport, Appendix 2 gives the equation in three dimensions, where a new circulatory term (term with non-zero curl) arises. Furthermore, although Fourier transforms in both spatial and temporal variables will be used throughout the development so as to deal with purely algebraic equations, the resulting expressions will be rational polynomials in the transformed variables, and as such the inverse transform may easily be taken to yield corresponding partial differential equations.

There is a subtlety when comparing the following work to the previous work of [8], touched upon in Sec. 2A. In the earlier work, a separate diffusive, high-frequency (HF) channel had to be explicitly invoked in order to derive the total heat-flux. The cut-off between HF and LF modes was set arbitrarily. In contrast, in this work, all modes contributing to transport are treated on an equal footing, thereby removing the need for an LF-HF cut-off, the only role of the HF channel being to establish a local temperature. This changes our classification of the EFL from "two-channel" to "quasi-full spectral".

The basis of the EFL is the assumption that quasi-ballistic modes do not interact with each other due to the small phase-space for such scattering [4], but can exchange energy with the reservoir, which is assumed to exist at temperature $T$. We denote the distribution function for quasi-ballistic modes as $g(x, \boldsymbol{k})$, where all spatial variation is assumed to be along the $x$-direction, and $\boldsymbol{k}$ is the phonon mode wave-vector of magnitude $k$ and making an angle $\theta$ with the $x$-axis.

We expand the distribution function in terms of spherical harmonics $P_l(cos\theta)$. Spherical harmonics form a complete, orthogonal basis for expanding angle-dependent azimuthally symmetric functions [16]:

$$g(x, \boldsymbol{k}, t) = \sum_{l=0}^{\infty} g_l(x, k, t) P_l(cos\theta) \qquad (1)$$

For the sake of brevity, we henceforth suppress the $(x, k, t)$ dependence wherever no ambiguity arises. Also, here and henceforth, symbols in bold fonts denote vectors, while the same symbols in normal fonts denote their magnitudes.

We first derive the equation for phonon modes of wave-vector magnitude $k$, lifetime $\tau(k)$, and group-velocity magnitude $v(k)$. The frequency of a quasi-ballistic mode of wave-vector $\boldsymbol{k}$ is denoted by $\omega(k)$. We also assume isotropic phonon dispersion, so that $\boldsymbol{v} = v\frac{\boldsymbol{k}}{k}$. The time-dependent linearized BTE for the quasi-ballistic modes is then given by

$$v cos\theta \frac{\partial g(x, \boldsymbol{k}, t)}{\partial x} + \frac{\partial g(x, \boldsymbol{k}, t)}{\partial t} = -\frac{g(x, \boldsymbol{k}, t) - f_{Eq}(x, k, T)}{\tau} \qquad (2)$$

We have earlier assumed the existence of a local temperature $T$ and thereby a corresponding spherically symmetric equilibrium distribution, familiar from Bose statistics: $f_{Eq}(x, \boldsymbol{k}, T) = f_{Eq}(x, k, T) \equiv f_{Eq}(T)$. This local temperature is established by some high-capacity reservoir. Eq. (2) is the quasi-ballistic-mode BTE in the absence of source terms; thus we assume that no external source of heat couples to the quasi-ballistic modes. We begin with the observation that, owing to the orthogonality of the spherical harmonics [16], the *x*-component of the quasi-ballistic heat-flux is determined solely by the first spherical harmonic $g_1$:

$$Q(x,t) = 2\pi \sum_k \int_{\theta=0}^{\pi} \hbar\omega g(x, \boldsymbol{k}, t) v\cos\theta\sin\theta d\theta = \frac{4\pi}{3} \sum_k \hbar\omega v g_1(x, k, t) \qquad (3)$$

This may be seen by substituting Eq. (1) for $g(x, \boldsymbol{k}, t)$ in the integrand and applying orthogonality of the spherical harmonics. Specifically, $cos\theta$ is the *l*=1 spherical harmonic, and since all other spherical harmonics are orthogonal to it, only the *l*=1 term survives.   Therefore we seek a differential equation for $g_1$. Here and henceforth, $\sum_k I(k)$ is shorthand for $\frac{1}{(2\pi)^3} \int dk I(k) k^2$ where $I(\boxed{?})$ is any function of *k,* and the integral is over all quasi-ballistic mode wave-vector magnitudes.

Highly non-equilibrium transport in electron gases has been investigated analytically by Baraff [17]; we adopt his approach. First we take the Fourier transform in time of Eq. (2), with transform variable $\gamma$. Substituting Eq. (1) into the Boltzmann transport equation, Eq. (2), multiplying successively by $P_{l'}(cos\theta)sin\theta$ for $l' = 0,1,2,...$ and integrating over $\theta$,  we arrive at a hierarchy of coupled equations for the $g_l$s, [17] the first three of which are

$$\frac{1}{3}v\frac{\partial g_1}{\partial x} + \frac{g_0}{\tilde{\tau}} - \frac{f_{Eq}(T)}{\tau} = 0 \qquad (4a)$$

$$\frac{2}{5}v\frac{\partial g_2}{\partial x} + v\frac{\partial g_0}{\partial x} + \frac{g_1}{\tilde{\tau}} = 0 \qquad (4b)$$

$$\frac{3}{7}v\frac{\partial g_3}{\partial x} + \frac{2}{3}v\frac{\partial g_1}{\partial x} + \frac{g_2}{\tilde{\tau}} = 0 \qquad (4c)$$

Here $\tilde{\tau}$ is given by $\frac{1}{\tilde{\tau}} = \frac{1}{\tau} + j\gamma$ where, as before, $\gamma$ is the Fourier transform variable in time, and $j$ is the imaginary unit. We truncate the hierarchy at the second order by setting $g_3 = 0$; other truncations are possible [17]. The generalization of Eqs. (4) to arbitrary order in spherical harmonics is given in Appendix 2. Therefore, this is the leading approximation beyond the Fourier law, which consists of setting $g_2 = 0$. Substituting Eq. (4c) into Eq. (4b) to eliminate $g_2$, and the result into Eq. (4a) to eliminate $g_0$, we arrive at an equation solely in terms of $g_1$:

$$-\frac{3}{5}(v\tau)^2\frac{\partial^2 g_1}{\partial x^2} + v\tau\frac{\partial f_{Eq}(T)}{\partial x} + (1+j\gamma\tau)^2 g_1 = 0 \qquad (5)$$

$f_{Eq}(T)$ depends on $x$ only through $T$, enabling the replacement $\frac{\partial f_{Eq}(T)}{\partial x} = \frac{\partial f_{Eq}(T)}{\partial T}\frac{\partial T}{\partial x}$.

We make the substitution here of $v(k)\tau(k) = \Lambda(k)$, the $k$- dependent MFP of quasi-ballistic phonons, to yield:

$$-\frac{3}{5}\Lambda^2\frac{\partial^2 g_1}{\partial x^2} + \Lambda\frac{\partial f_{Eq}(T)}{\partial T}\frac{\partial T}{\partial x} + (1+j\gamma\tau)^2 g_1 = 0 \qquad (6)$$

This is the point of departure from the original EFL. Taking the spatial Fourier transform $G_1(\chi, k, \gamma)$ of $g_1(x, k, \gamma)$ with transform variable $\chi$ replacing $x$, and rearranging, we get

$$G_1(\chi, k, \gamma) = \frac{-j\Lambda\frac{\partial f_{Eq}(T)}{\partial T}\chi T}{(1+j\gamma\tau)^2 + \frac{3}{5}\Lambda^2\chi^2} \qquad (7)$$

Using Eq. (3), we arrive at:

$$Q(\chi, \gamma) = -j\chi T\int_{k=0}^{k_{max}}\frac{\frac{1}{3}C(k)v(k)\Lambda(k)}{(1+j\gamma\tau)^2 + \frac{3}{5}\Lambda^2\chi^2}dk = -j\chi T\int_{k=0}^{k_{max}}\frac{\kappa_{diff}(k)}{(1+j\gamma\tau)^2 + \frac{3}{5}\Lambda^2\chi^2}dk \qquad (8)$$

where the differential heat-capacity with respect to $k$, $C(k) = \frac{1}{(2\pi)^3} 4\pi \hbar \omega(k) \frac{\partial f_{Eq}(T)}{\partial T} k^2$, and the differential thermal conductivity $\kappa_{diff}(k) = \frac{1}{3} C(k) v(k) \Lambda(k)$. Eq. (8) is equation of the generalized EFL for the Fourier transform of the net heat-flux.

## 4. CONNECTION TO THE QUASI-BALLISTIC TRANSPORT THEORY OF MAZNEV *ET AL* [4]

If we interpret $\chi$ as the inverse spatial period in the transient grating experiment, and restrict ourselves to such small phonon lifetimes that $\gamma\tau \ll 1$, it is seen that Eq. (8) is of the same form as Eq. (18) of Maznev *et al.* Specifically, the "correction factor" or "suppression function" of their work, $A_{Maznev}(\chi\Lambda) = \frac{3}{\chi^2\Lambda^2}\left(1 - \frac{arctan(\chi\Lambda)}{\chi\Lambda}\right)$ is replaced by $A_{EFL}(\chi\Lambda) = \frac{1}{1+\frac{3}{5}\Lambda^2\chi^2}$. Fig. 1 compares the two functions for a broad range of values of $\chi\Lambda$, and it is seen that they match very closely. Since the work of Maznev *et al.* treats the exact two-channel BTE, the truncation to the second order of our spherical harmonic expansion gives very small errors indeed.

Our formulation has the important advantage of yielding families of generalized Fourier laws of various finesse since $A_{EFL}(\chi\Lambda)$ is the ratio of rational polynomials. To wit, we use this formulation to generate a generalized EFL for a material whose MFP spectrum can be decomposed into a high-frequency (HF) channel and two low-frequency (LF) channels. To this end, consider a cut-off mean-free path $\Lambda_F = \Lambda(k_F)$ such that $\Lambda^2\chi^2 \ll 1$ for all length scales $1/\chi$ of interest and all $\Lambda < \Lambda_F$. For $\Lambda \geq \Lambda_F$, let the differential conductivity be given by

$$\kappa_{diff}(k) = \Delta\kappa_1 \delta(\Lambda - \Lambda_F) + \Delta\kappa_2 \delta(\Lambda - \Lambda_2) \tag{9}$$

$\delta$ is the Dirac delta function. Fig. 2 shows a schematic of the MFP accumulation function corresponding to this $\kappa_{diff}$.

Substituting Eq. (9) in Eq. (8) and simplifying, still under the assumption that $\gamma\tau \ll 1$, we get

$$\left(1 + \frac{3}{5}\Lambda_F^2\chi^2\right)\left(1 + \frac{3}{5}\Lambda_2^2\chi^2\right)(Q(\chi) + i\chi T\kappa_F) = \Delta\kappa_1(-j\chi T)\left(1 + \frac{3}{5}\Lambda_2^2\chi^2\right) + \Delta\kappa_2(-j\chi T)\left(1 + \frac{3}{5}\Lambda_F^2\chi^2\right)$$

(10)

where using the assumption $\Lambda^2\chi^2 \ll 1$, we have defined

$$\kappa_F = \int_{k=0}^{k_F} \frac{\kappa_{diff}(k)}{1 + \frac{3}{5}\Lambda^2\chi^2} dk \sim \int_{k=0}^{k_F} \kappa_{diff}(k)\, dk \tag{11}$$

We may easily take the spatial inverse Fourier transform of Eq. (10). Discarding fourth-order derivatives with respect to x, we get

$$Q(x) = \frac{3}{5}(\Lambda_2^2 + \Lambda_F^2)\frac{\partial^2 Q}{\partial x^2} - \kappa_{bulk}\frac{\partial T}{\partial x} + \frac{3}{5}\left[\Lambda_F^2(\kappa_{bulk} - \Delta\kappa_1) + \Lambda_2^2(\kappa_{bulk} - \Delta\kappa_2)\right]\frac{\partial^3 T}{\partial x^3} \tag{12}$$

where $\kappa_{bulk} = \kappa_F + \Delta\kappa_1 + \Delta\kappa_2$ is the bulk thermal conductivity, the sum of all phonon contributions. The generalization of Eq. (12) to an arbitrary number of quasi-ballistic channels may be performed along similar lines.

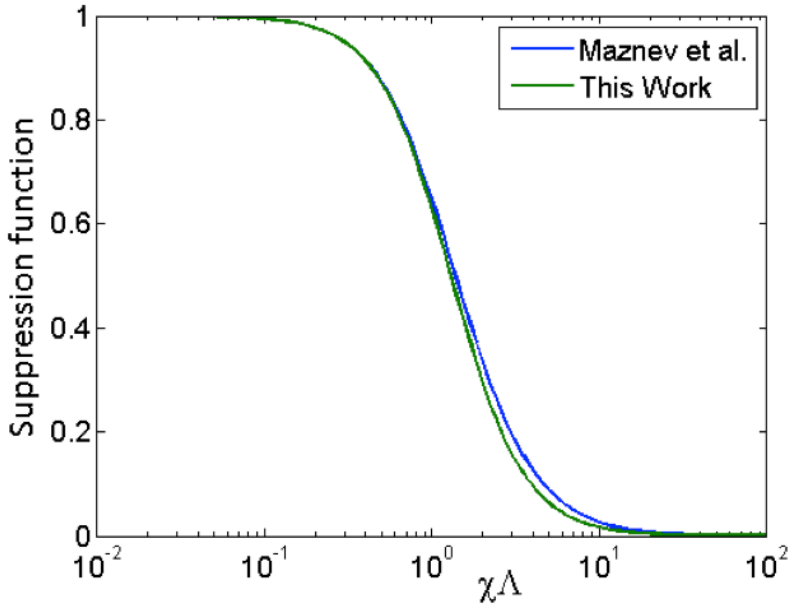

Fig. 1: Suppression functions of Maznev *et al.*[4] and this work (see Eq. (8)) match very closely over a large range of values of $\chi\Lambda$

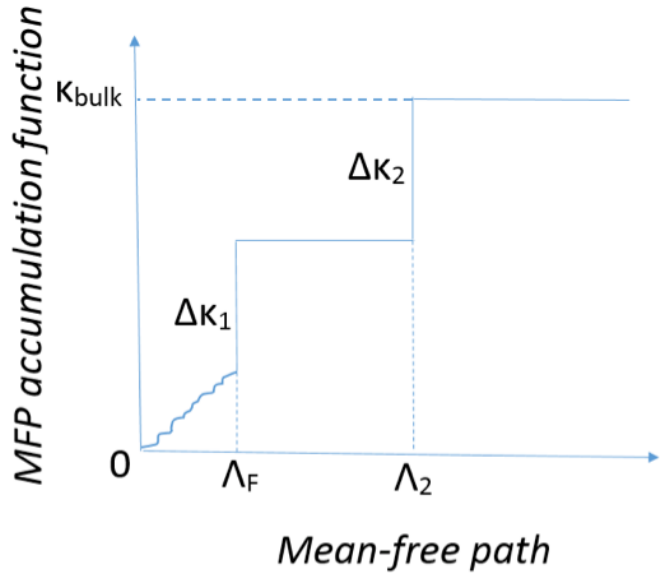

Fig. 2: Schematic of the mean-free path accumulation function corresponding to the differential conductivity of Eq. (9) used for deriving a three-channel enhanced Fourier law, Eq. (12).

## 5. CONTACT WITH CHEN'S BALLISTIC DIFFUSIVE EQUATIONS [6] AND MIRANDA'S *ET AL.*'S BTE SOLUTION [7]

Chen's ballistic-diffusive equations (BDE) and the solution of Miranda *et al.* are integral formulations of the Boltzmann transport equation, and as such incorporate explicitly the boundary and initial conditions, as opposed to differential formulations, such as the EFL that incorporate them indirectly. Due to this fundamental difference, only a few special consequences of each can be examined, namely those involving the diffusive part of the heat-flux. Using the EFL, we may recover an important feature of Miranda *et al.*, namely the form of the time-evolution operator $\left(1 + \tau \frac{\partial}{\partial t}\right)^2$ as opposed to the form $\left(1 + \tau \frac{\partial}{\partial t}\right)$ of the Cattaneo equation. The BDEs formulate the ballistic part of the heat flux again as the direct solution of an integral equation. In contrast to the EFL formulation, this approach is exact for highly out-of-equilibrium modes; however it suffers from the disadvantage of requiring a complicated formalism for their elucidation. Comparing the diffusive part of the BDE and EFL, we see that the latter has the time-evolution operator $\left(1 + \tau \frac{\partial}{\partial t}\right)^2$ while the BDE has the Cattaneo operator. Miranda *et al.* points out that invoking the square of the time-evolution operator may help solve the negative-temperature issue of the Cattaneo approach.

## 6. CONNECTION WITH THE TRANSPORT REGIMES OF HUA AND MINNICH [5]

Hua and Minnich have explored various regimes of heat conduction, from the ballistic, through the "strongly quasi-ballistic" and the "weakly quasi-ballistic" to the diffusive, using a semi-analytical solution of the spectrally resolved Boltzmann transport equation (see Sec. 2C). In Sec. 4 we have discussed in detail the near-equivalence of the EFL to the weakly quasi-ballistic limit. In the context of the transient grating experiment, in addition to the "Knudsen number" $\chi \Lambda$, Hua and Minnich define another dimensionless parameter $\eta(k) = \tau / \Gamma$ where $\tau(k)$ is the lifetime of modes of wave-vector magnitude $k$ and $\Gamma$ is the time-scale of transient grating decay. According to Eq. (8), for the $k$ component of the heat-flux $Q(k)$, the EFL gives the following equation:

$$Q(k) = -j\chi T \frac{\kappa_{diff}(k)}{(1 + j\gamma\tau)^2 + \frac{3}{5}\Lambda^2 \chi^2} \tag{13}$$

Setting the transform variable $-j\gamma$ to $1/\Gamma$ we arrive at the following equation:

$$\tilde{Q}(k) = -j\chi \tilde{T} \kappa_{diff}(k) \left[ A_{EFL}(\chi\Lambda) \frac{\left(1 + \frac{3}{5}\Lambda^2 \chi^2\right)}{(1 - \eta)^2 + \frac{3}{5}\Lambda^2 \chi^2} \right] \tag{14}$$

We have seen earlier (Sec. 4) that $A_{EFL}$ and $A_{Maznev}$ match very closely. The term within square brackets in Eq. (14) is the new suppression function. It is seen that it does not match the exact suppression function of Hua and Minnich $A_{HM}$ to any order in $\eta(k)$:

$$A_{HM} = A_{Maznev} (\chi\Lambda)\left(1 + \eta(k)\right) \tag{15}$$

The EFL corresponds to a second-order spherical harmonic expansion of the distribution function. Clearly by ignoring higher orders, it cannot describe strongly quasi-ballistic ballistic transport. Thus the EFL is not recommended for situations where the overall time constant of the experiment is expected to be on the order of the phonon lifetime.

## 7. A 2-OMEGA TECHNIQUE PROPOSED TO EXTRACT THE MEAN-FREE PATH SPECTRUM OF BULK MATTER

We propose a novel experiment to deduce the mean-free path accumulation function from purely electrical measurements. We propose to excite a metal line deposited on the material to be characterized with an alternating current with frequency in the range of 100-2000 Hz – please see Fig. 3. The AC voltage developed on the heater is measured across the inner pads of the heater. A temperature sensing line is deposited a short distance from the heater line. The distance between the heater and thermometer would be on the order of the mean-free path of dominant low-frequency modes (500 nm – 2000 nm for bulk Si and GaAs) [18]

The temperature of the thermometer line is extracted by passing a direct current and measuring the second harmonic voltage across its inner pads (Fig. 3). This is a variant of the AC third-harmonic technique originally invented by Atalla and coworkers [19] and popularized by

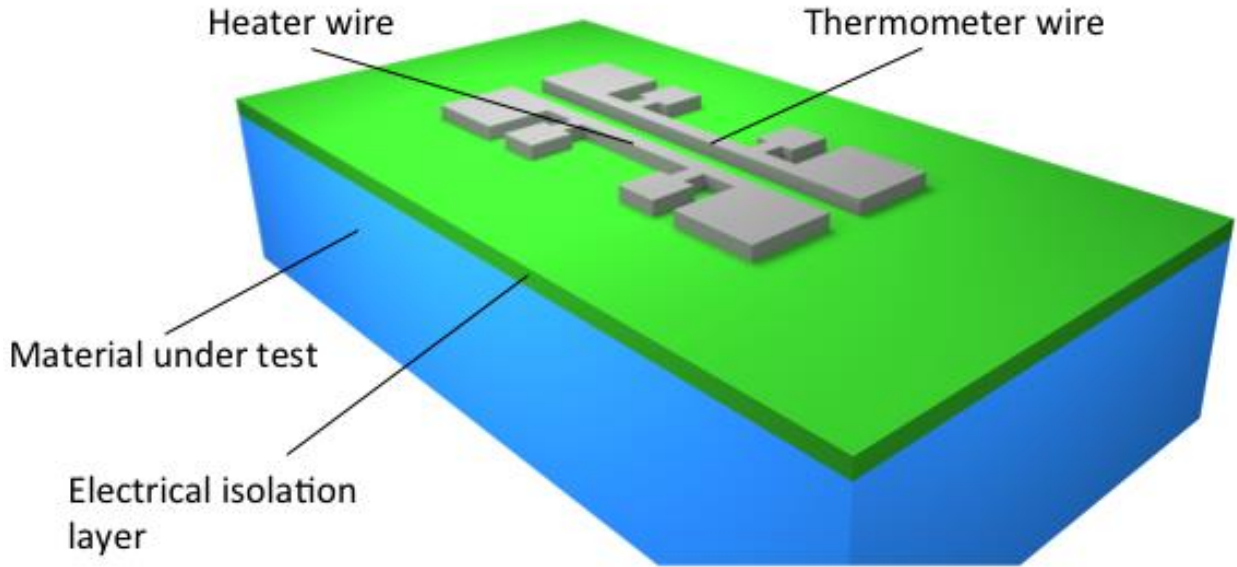

Fig. 3: Schematic of proposed experiment to investigate non-Fourier heat transfer

Cahill and coworkers [20] under the label "3-omega method". We note that our "2-omega" method was originally developed to measure the thermal anisotropy of bulk materials [21-22].

The three-dimension version of the EFL is derived in [23]. When generalized to an arbitrary phonon population, the equations read, assuming all variables to have an implicit time dependence $e^{-j2\omega t}$, and ignoring circulation of heat-flux (please see Appendix 2),

$$\boldsymbol{q}^{HF} = -\kappa^{HF}\nabla T \qquad (16)$$

$$\boldsymbol{q}_i^{LF} = -\kappa_i^{LF}\nabla T + \frac{3}{5}\Lambda_{\mathrm{i}}^2\nabla(\nabla \cdot \boldsymbol{q}_i^{LF}) \qquad (17)$$

Next we state the law of energy conservation:

$$\nabla \cdot (\boldsymbol{q}^{HF} + \textstyle\sum_i \boldsymbol{q}_i^{LF}) = -j2\omega C_v T \qquad (18)$$

$\boldsymbol{q}^{HF}$ is the high-frequency heat-flux that follows the Fourier law, $\kappa^{HF}$ is the corresponding thermal conductivity, $\boldsymbol{q}_i^{LF}$ is the low-frequency quasi-ballistic mode indexed by *i*, $\Lambda_{\mathrm{i}}$ is its mean-free path, $\kappa_i^{LF}$ is the kinetic theory value of the each LF mode's thermal conductivity, *T* is the temperature, $\omega$ is the angular frequency of the heater current and $C_v$ is the heat capacity of all phonon modes (HF and LF).

Eqs. (16)-(18) are less general than the formalism allows. Specifically, Eq. (16) assumes that some of the modes have small mean-free paths compared to the smallest length scale of the experiment and therefore travel diffusively. This 'lumping' of modes is often necessary in practice since the detailed mean-free path spectrum is not known at those length scales, and a cut-off wave-vector (or a cut-off mean-free path) must be invoked to separate the quasi-ballistic and diffusive modes. This cut-off mean-free path may be determined by fitting experimental data to a model with a specific number of quasi-ballistic channels indexed by $i$. For example, we show later for the 2-omega experiment that fitting a sample numerical dataset to a model with one quasi-ballistic channel (apart from the diffusive) yields the lumped thermal conductivity carried by modes with mean-free paths less than or equal to 0.37 times the experimental length-scale.

Equations (16-18) together with appropriate boundary conditions are solved in the appendix. Here we state the solution for the temperature sensed by the thermometer line of width $w_t$ separated from a heater of width $w_h$ by a center-to-center distance $D$.

$$T_{meas} = -\frac{P}{\pi \kappa^{bulk} L w_h w_t} \int_{x=-w_h/2}^{x=w_h/2} dx \int_{\tau=-w_t/2}^{\tau=w_t/2} d\tau \left( K_0\big(q(D + \tau - x)\big) + \sum_i \frac{\kappa_i^{LF}}{\kappa^{HF}} K_0\big(\epsilon_i(D + \tau - x)\big) \right)$$

(19)

Here $q = \sqrt{\frac{2j\omega C_v}{\kappa^{bulk}}}$ contains the only dependence on the heater current frequency, and $\kappa^{bulk} = \kappa^{HF} + \sum_i \kappa_i^{LF}$

In Fig. 4, we consider the mean-free path accumulation function of GaAs at room temperature measured by Freedman and co-workers [18] using frequency-domain thermoreflectance. This accumulation function may be approximated with four quasi-ballistic channels as shown. This yields the values for the parameters given in Table 1. Using these values for $\kappa^{HF}$, and $\Lambda_i$, $\kappa_i^{LF}$ for $i$ = 1-4, we get the measured temperature (in-phase or real component) versus heater frequency shown in Fig. 5 for heater/thermometer widths of 500 nm and center-to-center separation of 1000 nm. We also show the Fourier law result for comparison, demonstrating a substantial, frequency-independent deviation of the generalized EFL result from the Fourier law. This deviation is plotted as a function of heater-thermometer separation in Fig. 6. It is seen that beyond a separation of 3 microns, the deviation is practically zero, and thermal transport obeys the usual Fourier law.

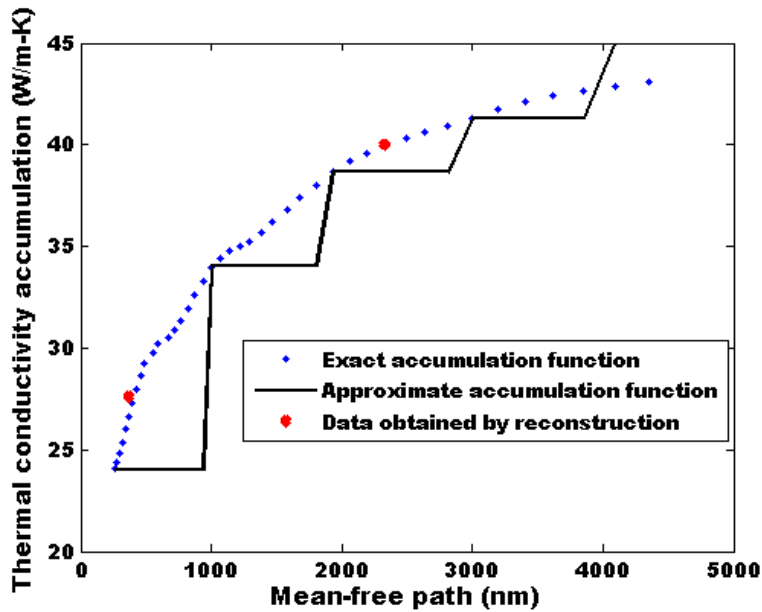

Fig. 4: The exact accumulation function (After Freedman *et al.* [18]), and the approximation used to generate the experimental predictions. Also shown is the result of a numerical experiment – please see text for details

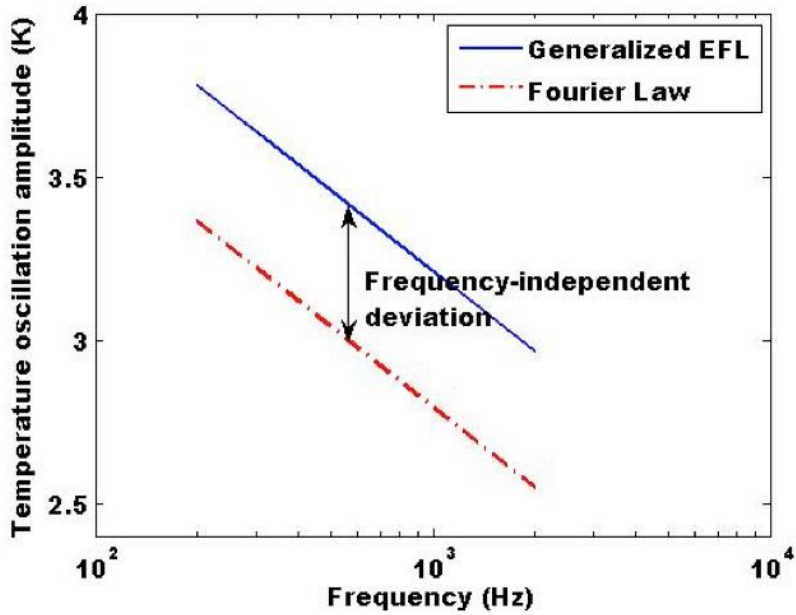

Fig. 5: Generalized EFL and Fourier law predictions versus frequency for heater-thermometer

separation 1000 nm and line widths 500 nm each. The input AC power is normalized to 1 W/cm.

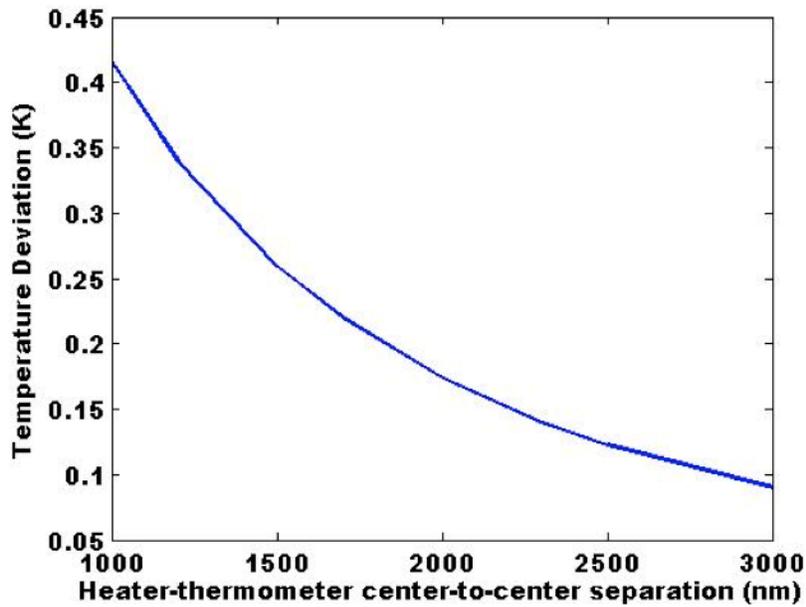

Fig. 6: Deviation from the Fourier law versus heater-thermometer separation. Heater and thermometer line-widths are 500 nm each. The input AC power is normalized to 1 W/cm. It is seen that the deviation tends to zero as the separation increases beyond 3000 nm.

We next look at the important inverse problem, i.e. how much information about the accumulation function may be obtained from given experimental data. For this purpose, we perform a numerical experiment as follows: we discretize Fig. 6 into four 'data' points, at heater-thermometer separations of 1000, 1500, 2000 and 2500 nm. We were able to fit this data satisfactorily (well within an assumed 10% error bar on the experiment) using only two channels, one following the Fourier law and one quasi-ballistic channel – please see Fig. 7. We obtained (a) $\kappa^{HF}$=26.7 W/m-K, namely the conductivity carried by modes obeying the Fourier law, and (b) $\Lambda_1$=2.3X10$^{-6}$ m, namely the mean-free path of the single quasi-ballistic channel. Now referring to the original accumulation function of Fig. 4, we see that these values demarcate the following points: (a) where the mean-free path is 0.37 times the smallest experimental length scale (1000 nm here), and (b) where the conductivity accumulates to about 88% of the bulk value.

This - (a) and (b) - is the maximum amount of data that may be reasonably obtained from the experiment, if four heater-thermometers separations are analyzed and the error bar is +/-5%. This is not a major limitation of the technique since it is expected to explore the several micron regime difficult to access by thermoreflectance. Moreover, we have noted that frequency-domain thermoreflectance [11] as well as transient gratings experiments [8] may similarly be explained by a compact two-parameter set.

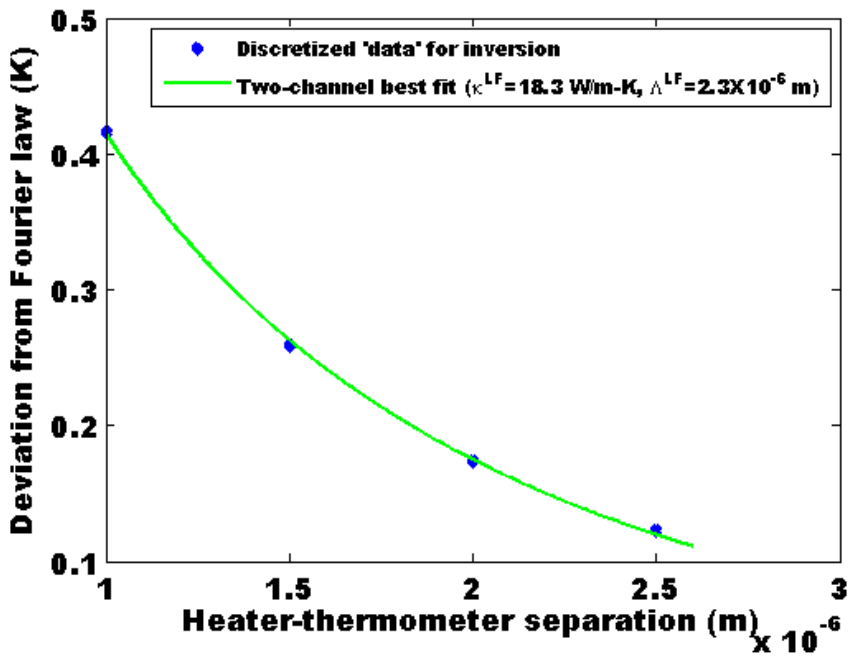

Fig. 7: The plot of Fig. 6 is discretized into 4 'data' points and fit satisfactorily to a two-channel model with the parameters shown. Please see text for details.

A few words are in order here on the significance of the proposed technique for measuring the mean-free path accumulation function. The most important advantage over time-domain or frequency-domain thermoreflectance is that interfacial impedances have very little impact on the measurement. This is because the heat-flux underneath the thermometer line is almost parallel to it. In practice, a 50 nm oxide insulating layer will be required to electrically isolate the metal lines from the material under test. Including even this rather low interfacial thermal conductance (<20 MW/m²-K) in the simulations did not show a discernible effect on the predicted temperatures at all considered frequencies. Another advantage is the purely electrical nature of the heating and temperature sensing which makes it compatible with a large array of temperature-controlled test chambers that lack optical ports.

We close this section with a discussion of the feasibility and limitations of this experiment. With ultraviolet lithography and familiar metal lift-off processes, one can manufacture devices with lines 500 nm wide and separated by 1000 nm (center-to-center). This separation sets the minimum length-scale over which quasi-ballistic transport may be observed in this experiment. Thus the dimensions assumed in this paper are realizable in practice. The large aspect ratio between the line width and length (1:600 at least) poses concerns of device yield.

Table 1

| Parameter | Value |
|---|---|
| $\kappa^{HF}$ | 24 W/m-K |
| $\kappa_1^{LF}, \Lambda_1$ | 10 W/m-K, 1000 nm |
| $\kappa_2^{LF}, \Lambda_2$ | 4.66 W/m-K, 1930 nm |
| $\kappa_3^{LF}, \Lambda_3$ | 2.66 W/m-K, 3000 nm |
| $\kappa_4^{LF}, \Lambda_4$ | 3.68 W/m-K, 4000 nm |
| $\kappa^{bulk}$ | 45 W/m-K |

## 8. CONCLUSIONS

The enhanced Fourier law has been derived and generalized from a gray phonon population to an arbitrary one. The resulting suppression function for the effective thermal conductivity in the transient grating experiment has been shown to approximate results of Maznev *et al.* in the weakly quasi-ballistic transport regime. The chief advantage of the EFL is seen to be its formulation in terms of observables like the heat-flux and temperature, akin to the Fourier law but rigorous enough to be capable of describing quasi-ballistic phonon transport. This feature is highly attractive in the context of simple explanations of quasi-ballistic transport experiments like the transient grating and frequency domain thermoreflectance experiments, and is likely to promote physically accurate device thermal simulations. A novel purely electrical probe of the mean-free path accumulation function has been proposed with notable advantages over traditional optical methods. Its feasibility has been demonstrated by analyzing the experiment using the equations developed in this work.

**ACKNOWLEDGMENT**

We wish to thank Professor Ali Shakouri (Purdue University, USA), and Professor Carl Meinhart, Ms. Nicole I. Halaszynski and Dr. Nathan Weitzner (University of California Santa Barbara) for helpful discussions, and Dr. Justin Freedman and Professor Jonathan Malen (Carnegie Mellon University, USA) for kindly providing us with the accumulation function of GaAs used in Fig. 4.

**FUNDING**

This work was funded by the National Science Foundation, USA under project number CMMI-1363207.

**NOMENCLATURE**

$A$                suppression function

$B$                radius of infinitesimal heater, m (see Fig. 7, appendix),

$C_v$                heat capacity of all phonon modes, J/kg K

$C(k)$                Differential heat-capacity with respect to $k$

| | |
|---|---|
| $D$ | center-to-center distance between the heater and thermometer lines, m |
| $F_i$ | Divergence of the low-frequency heat flux mode $i$ (see Eq. 25) |
| $f_{Eq}$ | Bose equilibrium distribution function |
| G | spatial Fourier transform of $g$ |
| $g(x, \boldsymbol{k})$ | distribution function for quasi-ballistic modes |
| $\hbar$ | Planck's constant, J s |
| $J$ | Imaginary unit |
| $K$ | modified Bessel function of the second kind |
| $\boldsymbol{K}$ | phonon mode wave-number, $\mathrm{m}^{-1}$ |
| $L$ | length of heater line, m |
| $L$ | angular momentum quantum number |
| $P$ | Power fluxed by the heater line |
| $P_l$ | Spherical harmonic of order $l$ |
| Q(x) | Net heat-flux |
| $q^{HF}, q^{LF}$ | The high-frequency and low-frequency mode heat-fluxes respectively |
| $T$ | temperature, K |
| $t$ | time, s |
| $w_h$ | heater line width, m |
| $w_t$ | thermometer line width, m |
| $x, y, z$ | coordinates, m |
| $\alpha_i$ | constant for Eq. 29 |
| $\Gamma$ | Time-scale of transient grating decay |
| $\delta$ | Dirac delta function |
| $\eta$ | Ratio of lifetime of phonon mode to the transient grating decay time-scale |
| $\theta$ | angle of wave-vector with respect to transport axis |
| $\kappa$ | thermal conductivity, W/m K |
| $\Lambda_i$ | mean-free path of LF phonon mode indexed by '$i$' |
| $\tau_i(k)$ | lifetime of a mode indexed by '$i$' with wave-vector magnitude $k$, s |
| $v$ | group-velocity magnitude |
| $\chi$ | transformation of '$x$' coordinate variable, m |
| $\omega$ | angular frequency |
| $\nabla$ | gradient operator |
| $\epsilon_i$ | Defined by the equation $\frac{1}{\epsilon_i^2} = \frac{3}{5}\Lambda_i^2$ for each low-frequency phonon mode $i$ |
| AC | Alternating Current |
| BDE | Ballistic-Diffusive equations |
| BTE | Boltzmann transport equation |

A generalized enhanced Fourier law and underlying connections to major frameworks for quasi-ballistic phonon transport

| DC | Direct Current |
| EFL | Enhanced Fourier Law |
| Hz | Hertz |
| HF | High Frequency |
| LF | Low Frequency |
| MFP | Mean Free Path |

**APPENDIX 1**

In order to analyze the 2-omega experiment for exploring the phonon mean-free path spectrum, we derive the temperature profile on a thermometer of width $w_t$ separated from a heater of width $w_h$ by a center-to-center distance of $D$. All variables are assumed to have an implicit time dependence $e^{-j2\omega t}$.

First we state the enhanced Fourier law in three dimensions [23], modified to account for an arbitrary number of low-frequency channels

$$\boldsymbol{q}^{HF} = -\kappa^{HF}\nabla T \qquad (20)$$

$$\boldsymbol{q}_i^{LF} = -\kappa_i^{LF}\nabla T + \frac{3}{5}\Lambda_i^2 \nabla(\nabla \cdot \boldsymbol{q}_i^{LF}) \qquad (21)$$

Next we state the law of energy conservation:

$$\nabla \cdot (\boldsymbol{q}^{HF} + \sum_i \boldsymbol{q}_i^{LF}) = -j2\omega C_v T \tag{22}$$

$\boldsymbol{q}^{HF}$ is the high-frequency heat-flux that follows the Fourier law, $\kappa^{HF}$ is the corresponding thermal conductivity, $\boldsymbol{q}_i^{LF}$ is the low-frequency quasi-ballistic mode indexed by $i$, $\Lambda_i$ is its mean-free path, $\kappa_i^{LF}$ is the kinetic theory value of the modal thermal conductivity, $T$ is the temperature, $\omega$ is the angular frequency of the heater current and $C_v$ is the heat capacity of all phonon modes (HF and LF).

The boundary conditions are given by,

$$\boldsymbol{q}^{HF} \cdot \widehat{\boldsymbol{n}} = -P \tag{23}$$

$$\boldsymbol{q}^{LF} \cdot \widehat{\boldsymbol{n}} = 0 \tag{24}$$

Where $P$ is the input power from the infinitesimal heater

$$\nabla \cdot \boldsymbol{q}_i^{LF} = F_i \tag{25}$$

From (21)

$$\nabla^2 F_i - \epsilon_i^2 F_i = \kappa_i \epsilon_i^2 \nabla^2 T \tag{26}$$

Here $\frac{1}{\epsilon_i^2} = \frac{3}{5}\Lambda_i^2$. The solution can be written as a sum of local and boundary contributions:

$$F_i = F_{il} + F_{ib} \tag{27}$$

To leading order, $F_{il} = -\kappa_i \nabla^2 T \tag{28}$

$F_{ib}$ satisfies $\nabla^2 F_{ib} - \epsilon_i^2 F_{ib} = 0$ whose general solution that does not blow up far away from the heater is,

$$F_{ib} = \alpha_i K_0(\epsilon_i r) \tag{29}$$

Here, $K_0$ is the modified Bessel function of the second kind and $\alpha_i$ is a constant to be determined. To that end, consider an infinitesimally wide, very long (length = $L$) heater delivering a power $P$. Consider a semi-cylindrical region of radius $b$ around it, as shown in Fig. 8

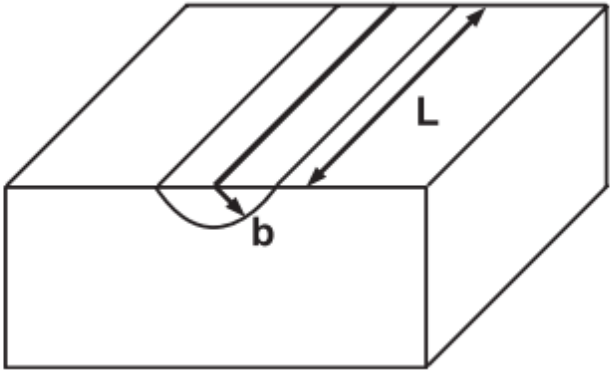

Fig. 8: Schematic of geometry used to derive the constant of integration in Eq. (29)

At the boundary, $\nabla \cdot \boldsymbol{q}_i^{LF} = F_{ib}$ by definition. From Eqs. (21), (24), (23) and (20), at the semi-cylindrical surface,

$$\nabla F_{ib} \cdot \widehat{e_r} = -\frac{\epsilon_i^2 \kappa_i^{LF}}{\kappa^{HF}} \frac{P}{\pi b L} \tag{30}$$

Here $\widehat{e_r}$ is the radial unit vector at the semi-cylindrical surface. Substituting in (29),

$$\alpha_i = -\frac{\epsilon_i^2 \kappa_i^{LF}}{\kappa^{HF}} \frac{P}{\pi L} \tag{31}$$

From (27), (28), (29) and (31) we have

$$F_i = -\kappa_i \nabla^2 T - \frac{\epsilon_i^2 \kappa_i^{LF}}{\kappa^{HF}} \frac{P}{\pi L} K_0(\epsilon_i r) \tag{32}$$

Using Eq. (22), (20) and (32),

$$\nabla^2 T - q^2 T = -\sum_i \frac{\epsilon_i^2 \kappa_i^{LF}}{\kappa^{HF} \kappa^{bulk}} \frac{P}{\pi L} K_0(\epsilon_i r) \tag{33}$$

Where $q = \sqrt{\frac{2j\omega C_v}{\kappa^{bulk}}}$ and $\kappa^{bulk} = \kappa^{HF} + \sum_i \kappa_i^{LF}$

The general solution to (33) is given by

$$T = C K_0(qr) - \sum_i \frac{\epsilon_i^2}{\epsilon_i^2 - q^2} \frac{\kappa_i^{LF}}{\kappa^{bulk}} \frac{P}{\pi \kappa^{HF} L} K_0(\epsilon_i r) \tag{34}$$

$C$ is a constant to be determined. Using boundary condition (23), and using $\epsilon_i^2 - q^2 \gg 0$ since the thermal penetration depth is much larger than the mean-free path in the frequency range of interest,

$$T \sim -\frac{P}{\pi \kappa^{bulk} L} K_0(qr) - \frac{P}{\pi \kappa^{bulk} L} \sum_i \frac{\kappa_i^{LF}}{\kappa^{HF}} K_0(\epsilon_i r) \tag{35}$$

This is the response everywhere due to an infinitesimally narrow heater. Convolving over the heater width and averaging over the thermometer width, with $D$ as the heater-thermometer separation,

$$T = -\frac{P}{\pi \kappa^{bulk} L w_h w_t} \int_{x=-w_h/2}^{x=w_h/2} dx \int_{\tau=-w_t/2}^{\tau=w_t/2} d\tau \left( K_0\big(q(D + \tau - x)\big) + \sum_i \frac{\kappa_i^{LF}}{\kappa^{HF}} K_0\big(\epsilon_i(D + \tau - x)\big) \right)$$

(36)

## APPENDIX 2

Here we suggest ways of generalizing the one-dimensional analysis, and generalize the EFL to three spatial dimensions.

Although we have truncated the spherical harmonic expansion of the distribution function at the $l$=2 order, it is possible to generalize to arbitrary order. Eqs. (4a-c) are part of a hierarchy of equations, the $l^{th}$ equation of which is, for $l$>0 [17]:

$$\frac{l+1}{2l+3} v \frac{\partial g_{l+1}}{\partial x} + \frac{l}{2l+1} v \frac{\partial g_{l-1}}{\partial x} + \frac{g_l}{\bar{\tau}} = 0 \qquad (37)$$

Of course, increasing the number of equations in the hierarchy improves the accuracy of the model. However this will result in a differential equation of higher order, requiring more boundary conditions than can be deduced from the physics of the problem.

In three dimensions, the equations of the generalized EFL and energy conservation take the following form [23], where all symbol connotations are the same as in Eqs (16)-(18) and $\tau_i$ is the lifetime of phonons in the $i$th channel.

$$\left(1 + \tau_i \frac{\partial}{\partial t}\right)^2 \boldsymbol{q}_i^{LF} = -\kappa_i^{LF} \nabla T + \frac{3}{5}(\Lambda_i^{LF})^2 \nabla(\nabla \cdot \boldsymbol{q}_i^{LF}) - \frac{1}{5}(\Lambda_i^{LF})^2 \nabla \times (\nabla \times \boldsymbol{q}_i^{LF}) \qquad (38)$$

$$\nabla \cdot (\boldsymbol{q}^{HF} + \sum_i \boldsymbol{q}_i^{LF}) = -C_v \frac{\partial T}{\partial t} \qquad (39)$$

Although Ref. [23] derived this equation starting from the steady-state BTE, extension to time-dependent BTE is simple using Fourier transforms, as illustrated in Sec. 3. The last term of Eq. (38) is a circulatory term, which we have ignored in Sec. 7. Inclusion of this term requires knowledge of the tangential heat-flux at the surface, which is not available except for white, specular boundaries. The physics of this term has only recently been touched upon [24] and we will not concern ourselves with it in this paper.

## Figure Captions List

Fig. 1    Suppression functions of Maznev *et al.*[4] and this work (see Eq. (8)) match very closely over a large range of values of $\chi\Lambda$

Fig. 2    Schematic of the mean-free path accumulation function corresponding to the differential conductivity of Eq. (9) used for deriving a three-channel enhanced Fourier law, Eq. (12)

Fig. 3    Schematic of proposed experiment to investigate non-Fourier heat transfer

Fig. 4    The exact accumulation function (After Freedman *et al.* [18]), and the approximation used to generate the experimental predictions

Fig. 5    Generalized EFL and Fourier law predictions vs frequency for heater-thermometer separation 1000 nm and line widths 500 nm each. The input AC power is normalized to 1 W/cm.

Fig. 6    Deviation from the Fourier law vs heater-thermometer separation. Heater and thermometer line-widths are 500 nm each. The input AC power is normalized to 1 W/cm. It is seen that the deviation tends to zero as the separation increases beyond 3000 nm.

Fig. 7:    Schematic of geometry used to derive the constant of integration in Eq. (29)

Fig. 8:    The plot of Fig. 6 is discretized into 4 'data' points and fit satisfactorily to a two-channel model with the parameters shown. Please see text for details.

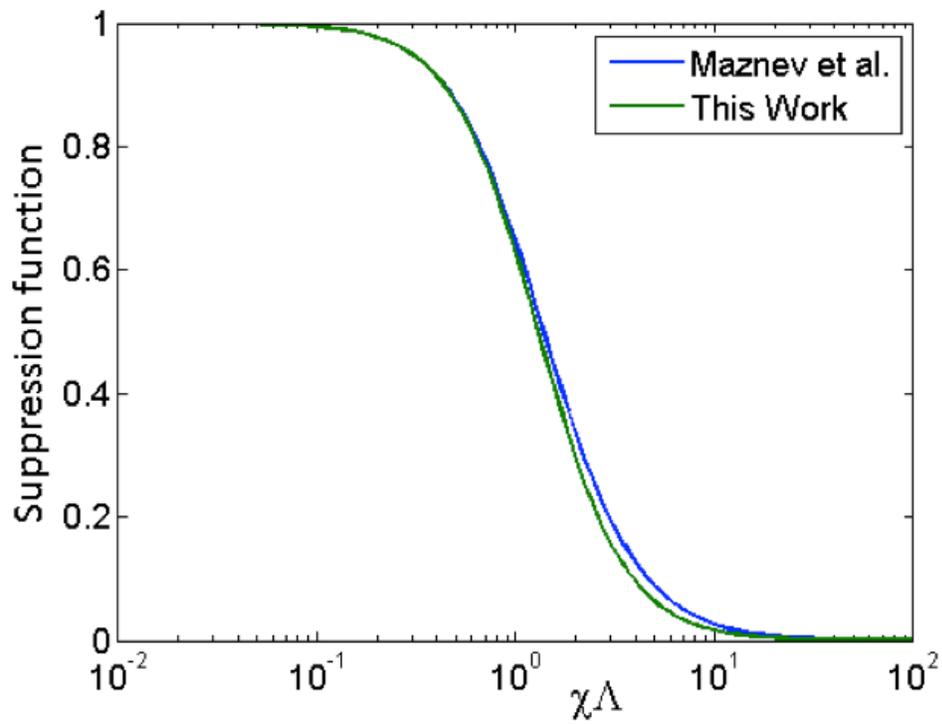

Fig. 1: Suppression functions of Maznev *et al.*[4] and this work (see Eq. (8)) match very closely over a large range of values of $\chi\Lambda$

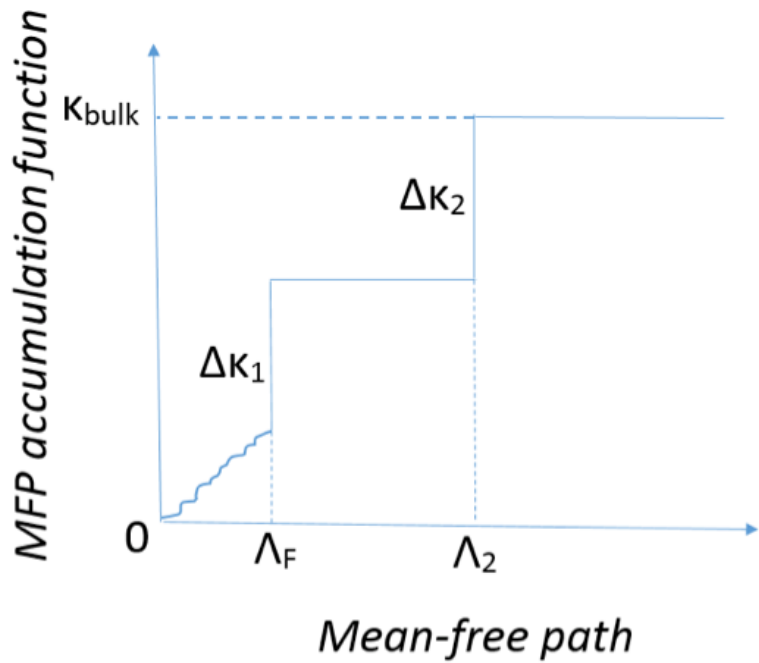

Fig. 2: Schematic of the mean-free path accumulation function corresponding to the differential conductivity of Eq. (9) used for deriving a three-channel enhanced Fourier law, Eq. (12)

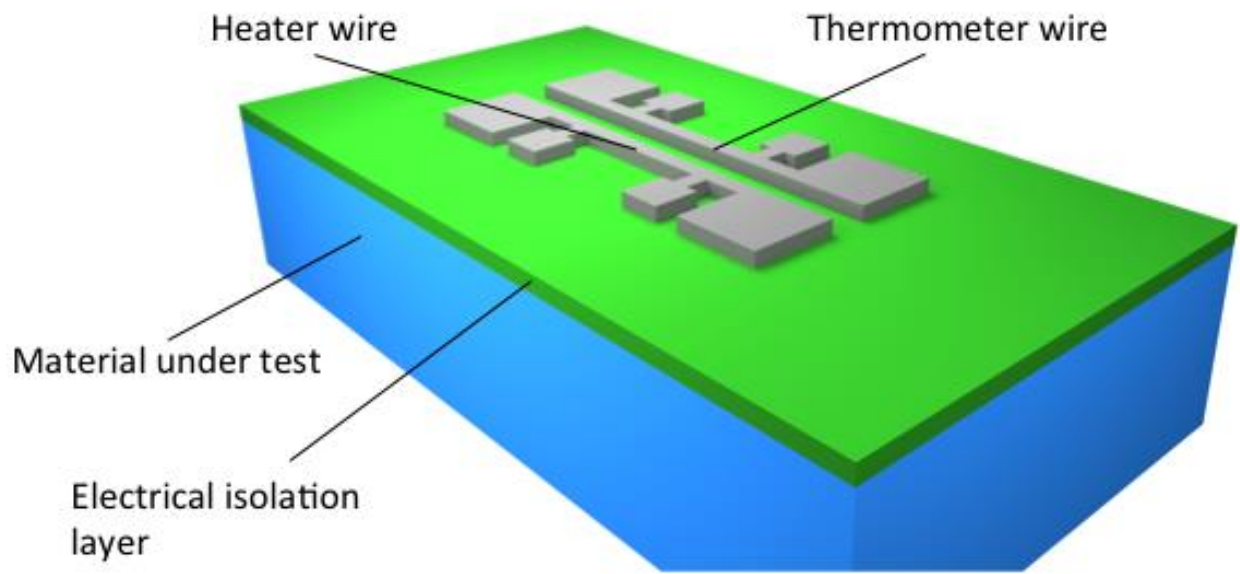

Fig. 3: Schematic of proposed experiment to investigate non-Fourier heat transfer

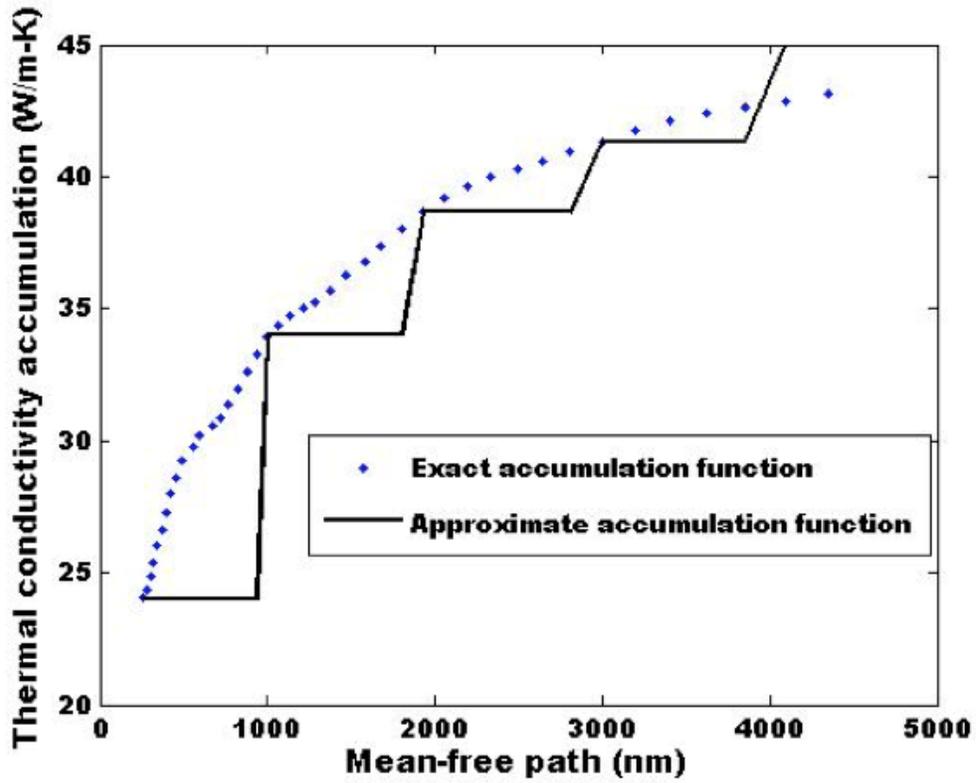

Fig. 4: The exact accumulation function (After Freedman *et al.* [18]), and the approximation used to generate the experimental predictions

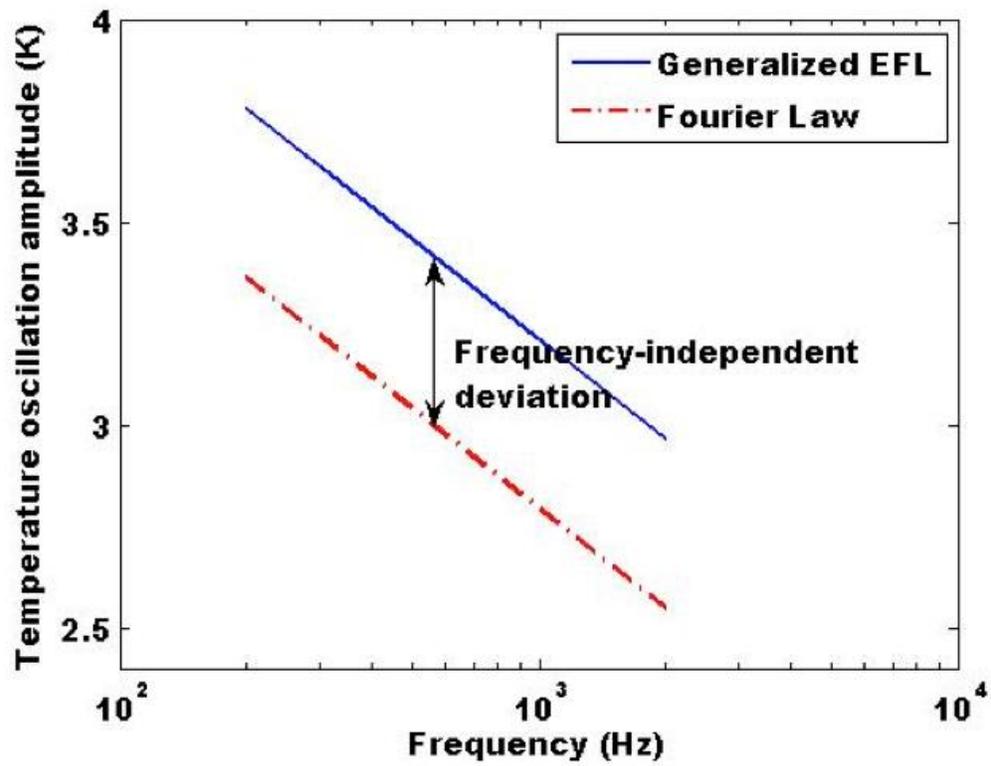

Fig. 5: Generalized EFL and Fourier law predictions vs frequency for heater-thermometer separation 1000 nm and line widths 500 nm each. The input AC power is normalized to 1 W/cm.

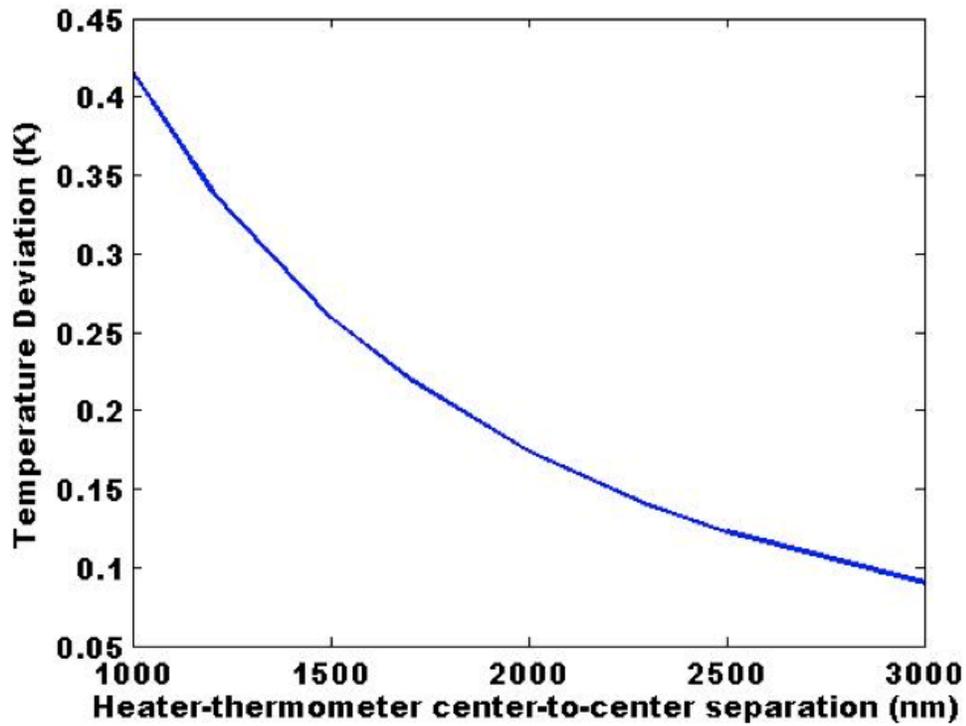

Fig. 6: Deviation from the Fourier law vs heater-thermometer separation. Heater and thermometer line-widths are 500 nm each. The input AC power is normalized to 1 W/cm. It is seen that the deviation tends to zero as the separation increases beyond 3000 nm.

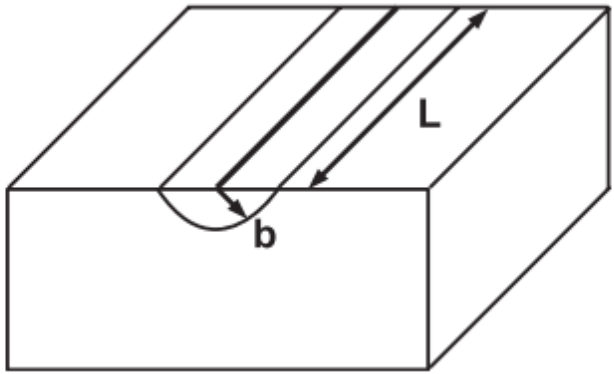

Fig. 7: Schematic of geometry used to derive the constant of integration in Eq. (29)

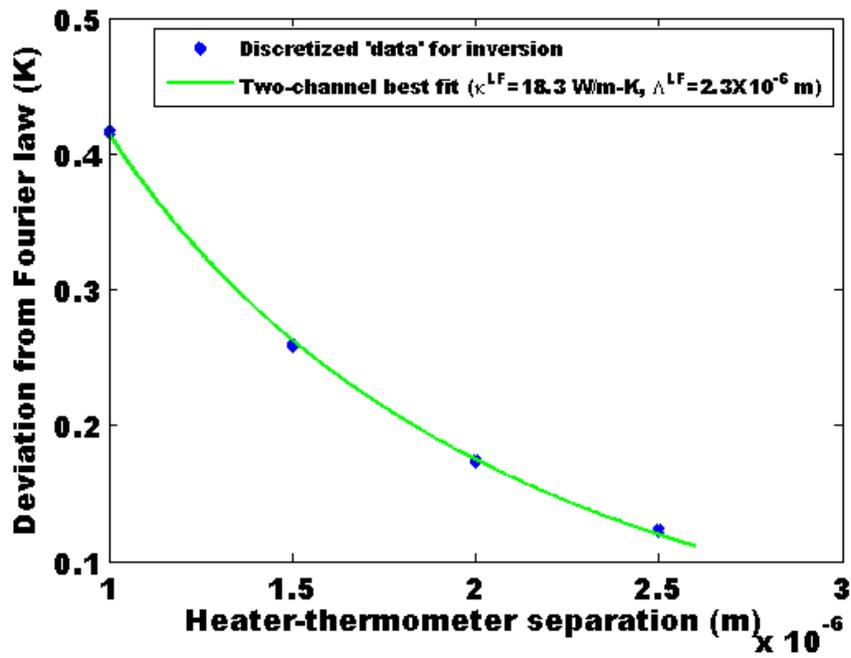

Fig. 8: The plot of Fig. 6 is discretized into 4 'data' points and fit satisfactorily to a two-channel model with the parameters shown. Please see text for details.